\documentclass[aps,amssymb,showpacs,pra,twocolumn,floatfix]{revtex4}

\usepackage{epsf}
\usepackage{amsmath}
\usepackage{graphicx}
                                                                                                                                  
\begin{document}
         
\title {Fluctuation interactions of colloidal particles}

\author{T. Ocampo-Delgado$^{1}$ and B. Ivlev$^{1,2}$}

\affiliation
{$^{1}$Instituto de F\'{\i}sica, Universidad Aut\'onoma de San Luis Potos\'{\i}
San Luis Potos\'{\i}, San Luis Potos\'{\i} 78000 Mexico\\
and\\
$^{2}$Department of Physics and Astronomy and NanoCenter,
University of South Carolina, Columbia, South Carolina 29208, USA}


\begin{abstract}

For like charged colloidal particles two mechanisms of attraction between them survive when the interparticle distance is larger than the 
Debye screening length. One of them is the conventional van der Waals attraction and the second one is the attraction mechanism mediated
by thermal fluctuations of particle position. The latter is related to an effective variable mass (Euler mass) of the particles 
produced by a fluid motion involved. The most stronger attraction potential (up to the value of temperature $T$) corresponds to a situation 
of uncharged particles and a relatively large Debye screening length. In this case the third mechanism of attraction enters the game. It is 
mediated by thermal fluctuations of fluid density. 

\end{abstract} \vskip 1.0cm

\pacs{05.20.Jj, 82.70.Dd}

\maketitle

\section{THE SHORT STORY}
\label{intr}
Systems of charged colloidal particles exhibit a variety of unusual physical properties \cite{DER,ISR,HEI}. Colloidal particles can be 
arranged into crystal \cite{PIE} and into structures with clusters and voids \cite{ONO,ITO,MEJ}. Colloidal systems may undergo different 
types of phases transition \cite{CHA,MON,LAR,RAD,LIN,SOW}. Topological phase transitions in two-dimensional system of colloidal particles 
have been discussed in \cite{MUR,ZAH}. Unusual ensembles of colloidal particles were observed in Refs.~\cite{JAIME1,JAIME2,JAIME3}. In 
\cite{CHO} buckling instabilities in confined colloidal crystal were analyzed. Interesting behaviors of colloidal in external fields were 
reported in \cite{RUB}. Colloidal particles accept in an electrolyte some surface charge, screened by counterions at Debye's length 
$\lambda_{D}$, which results in the repulsion potential of Derjaguin, Landau, Verwey, and Overbeek (DLVO) \cite{DER,ISR}. The DLVO theory, 
as a result of solution of linearized Poisson-Boltzmann equation, has been questioned in \cite{SIR,LOW}. The generalization of DLVO 
interaction via a modification of counterion screening was reported in \cite{OSP}.

The long range attraction of like charged particles is a matter of challenge and controversy in colloidal science. Very schematically, the 
story of the long range attraction is the following.

Initially, the long range attraction of colloidal particles on the micron scale has been reported in Ref.~\cite{FRADEN}. The authors 
reconstructed a pair potential from measured correlation functions of a large ensemble of colloidal particles.

The authors of Refs.~\cite{GREER1,GREER2,GREER3} published even more surprising results when the attraction was extended over a distance 
of almost 4~$\mu$m. They used a laser technique releasing two particles and observing their behavior. It happened that their measurements 
were misinterpreted. It was observed just a macroscopic hydromechanical effect associated with specificity of measurements but not a 
microscopic mechanism of attraction \cite{BRENNER}.

Bechinger and his group shown for a high concentrated particle ensemble that different types of interparticle potential
result in almost identical correlation functions \cite{BECH1,BECH2}. The conclusion was that an interaction potential should be 
drawn from experiments with a pair of colloidal particles but not from a large ensemble of them. Another difficulty is connected 
to an uncertainty in observed particle position caused by diffraction which can result in errors in calculated pair potential. See 
Ref.~\cite{MARKUS} and references therein. To reduce the diffraction uncertainty one should use ultraviolet observations. So reliability 
of calculation of pair potential on the basis of statistical properties of large ensembles of particles is questionable. 

Despite a lack of reliability of various experiments on determination of a microscopic mechanism of attraction of like charges it is always
intriguing whether they attract or not in reality. We consider the following microscopic mechanism of attraction of two colloidal particles
or a particle and a wall.

1. The conventional van der Waals attraction $u_{vdW}$ \cite{DZYAL} mediated by electromagnetic fluctuations.

2. The attraction $U_{com}$, mediated by thermal compression fluctuations of a fluid. 

3. The attraction $I$ associated with thermal fluctuations of particle positions in a fluid. It results from variable particle masses (Euler
masses) depending on a distance between them. This mechanism and the term ``Euler mass'' was proposed in Ref.~\cite{IVLEV}.

In the first mechanism an energy of fluctuating electromagnetic waves depends on a distance between particles and, therefore, it leads to a
force.

The second mechanism is generic with the first one since electromagnetic waves are just substituted by hydrodynamic ones.

In the third mechanism moving particles drag a part of a fluid. A mass of the involved fluid depends on a distance between the particles. 
There is a thermal drift of the particles into the region with a larger effective mass which is analogous to classical mechanics. This can 
be interpreted as an effective interaction mediated by thermal fluctuations of particle positions in a fluid. In formation of the variable 
mass high frequency fluctuations of the particles are involved when dissipative hydrodynamic effects are not important \cite{HYDR}. This 
corresponds to Euler hydrodynamics and gives rise to the term ``Euler mass''. The formation of Euler mass reminds the equipartition law
when a mean kinetic energy is $T/2$ regardless of dissipation. This is due to that the mean kinetic energy is also determined by high 
frequencies. 

There is a substantial difference among the above attraction mechanisms. The first and the third ones survive when an interparticle
distance becomes larger than the Debye screening length. On this distance the DLVO repulsion is very small and the above two mechanisms
are the only interaction. The second mechanism works when Coulomb effects in the fluid are not pronounced. Namely, the particles are not 
charged and the Debye screening length is larger than the interparticle distance. A relative role of attraction $I$ and $u_{vdW}$ was also
analyzed in Ref.~\cite{WID}.

A goal of this paper is to study the three above attraction potentials for two particles and for one near a wall (walls). 
\section{VAN DER WAALS INTERACTION}
\label{vdW}
The energy of fluctuating electromagnetic field around two particles depends on a distance $R$ between them and, therefore,
results in an interaction force. It is called van der Waals force \cite{DZYAL}. This force is mainly determined by a typical wave 
length $\lambda$ of the fluctuating electromagnetic field since it should be $\lambda\sim R$. The permittivity of the 
particles material $\varepsilon(\omega)$ depends on the typically frequency $\omega\sim c/R$. We consider relatively large 
interparticle distance
\begin{equation}
\label{1}
\frac{c}{\omega_0}<R,
\end{equation}
where $\omega_{0}$ corresponds to an absorption peak of $\varepsilon(\omega)$. For example, for water 
$\omega_{0}\sim 10^{16}~{\rm sec}^{-1}$. The estimate (\ref{1}) reads $100~\AA <R$.

At a finite temperature the typical wave length of the fluctuating electromagnetic field is $\hbar c/T$. Below the interparticle distance is
not too large
\begin{equation}
\label{2}
R<\frac{\hbar c}{T},
\end{equation}
which is equivalent to $R<7.4~\mu m$ at room temperature.

We consider a typical interparticle distance $R$ of the order of one or two microns which relates to the conditions (\ref{1}) and (\ref{2}).
In the optical interval of $\omega$ the permittivity is determined by a refractive index and can be substituted by a dielectric constant 
$\varepsilon$ for particles and a dielectric constant $\varepsilon_0$ for a surrounded medium. When the two dielectric constants are close
to each other
\begin{equation}
\label{3}
\frac{\varepsilon-\varepsilon_0}{\varepsilon_0}<<1
\end{equation}
one can use the approach of pairwise summation to calculate the energy of electromagnetic fluctuations (the van der Waals 
interaction energy) \cite{DZYAL,NIN1}
\begin{equation}
\label{4}
u_{vdW}(R)=-\frac{23\hbar c(\varepsilon-\varepsilon_0)^2}{64\pi^3\varepsilon_0^{5/2}}
\int_{V_1}d^3r_1\int_{V_2}\frac{d^3r_1}{|\vec{r}_1-\vec{r}_2|^7}.
\end{equation}

In Eq.~(\ref{4}) the integrations occur inside the volumes of two bodies. For two identical spherical particles of the radius $a$ and 
center-to-center distance $R$ the integration in Eq.~(\ref{4}) results in \cite{DZYAL,NIN1}
\begin{eqnarray}
\label{5}
u_{vdW}(R)=-\frac{23}{1920\pi}\frac{(\varepsilon-\varepsilon_0)^2}{\varepsilon_0^{5/2}}\frac{\hbar c}{R}
\bigg[\frac{2a^2(20a^2-3R^2)}{(R^2-4a^2)^2}\\
\nonumber
+\frac{2a^2}{R^2}+\ln{\frac{R^2}{R^2-4a^2}}\bigg]
\end{eqnarray}

For a spherical particle near a flat infinite wall (the both with the dielectric constant $\varepsilon$) 
Eq.~(\ref{4}) yields 
\begin{equation}
\label{6}
u_{vdW}(h)=-\frac{23}{640\pi}\frac{(\varepsilon-\varepsilon_0)^2}{\varepsilon_0^{5/2}}\frac{\hbar c}{a}
\int^{1}_{-1}\frac{(1-z^2)dz}{(z+h/a)^4}.
\end{equation}
In Eq.~(\ref{6}) $h$ is the center-to-wall distance.

In the region of visible light for water $\varepsilon_{0}\simeq 1.77$ and for usually used polystyrene colloidal particles 
$\varepsilon\simeq 2.40$. So the parameter $(\varepsilon-\varepsilon_{0})/\varepsilon_{0}\simeq 0.35$ can be considered as relatively small 
and, therefore, the equations (\ref{5}) and (\ref{6}) are reasonable approximations for van der Waals interaction.

\section{INTERACTION MEDIATED BY COMPRESSION FLUCTUATIONS OF THE FLUID}
\label{comp}
Suppose two particles to be totally fixed inside a hydrodynamic medium and they serve only as obstacles for a fluid motion.
There is no macroscopic motion in the system and the only motion is caused by thermal 
fluctuations of the fluid velocity $\vec v(\vec r,t)$. In this case the free energy of thermal fluctuations of the fluid 
$F(R)$ depends on the distance $R$ between the particles. The function
\begin{equation}
\label{7}
U_{com}(R)=F(R)-F(\infty)
\end{equation}
is an interaction mediated by compression fluctuations of the fluid analogously to the conventional van der Waals interaction mediated 
by electromagnetic fluctuations. To find the free energy of thermal fluctuations of the fluid one can start with the linearized 
Navier-Stokes equation 
\cite{HYDR}
\begin{equation}
\label{8}
\rho\frac{\partial\vec v}{\partial t}=-\nabla p +\eta\nabla^{2}\vec v +\left(\zeta + 
\frac{\eta}{3}\right)\nabla{\rm div}\vec v.
\end{equation}
There are two types of fluid motion, one of them is a transverse diffusion and the second one is longitudinal sound waves associated with 
the density variation. The equilibrium free energy of transverse motions is determined by the Boltzmann distribution of their kinetic 
energies and does not depend on the friction coefficient in the thermal limit. The free energy of transverse motions in the thermal limit 
depends on the total volume, but not on relative positions of bodies. Therefore, transverse fluctuations do not result in an interaction. 

Quite opposite situation occurs for longitudinal motions, when the total free energy is a sum of energies of different 
sound modes. The spectrum of sound waves depends on the distance between bodies $R$ due to hydrodynamic boundary conditions
on body surfaces and this results in $R$-dependence of the free energy. Hence, the fluctuation interaction between bodies 
is mediated by hydrodynamic sound waves like the conventional van der Waals interaction is mediated by fluctuations of 
electromagnetic ones. Putting $\vec v=\nabla\partial\phi/\partial t$, one can obtain from Eq.~(\ref{8})
\begin{equation}
\label{9}
\rho\hspace{0.1cm}\frac{{\partial}^{2}\phi}{\partial t^{2}}=-\delta p + 
\left(\zeta + \frac{4\eta}{3}\right)\frac{\partial}{\partial t}\nabla^{2}\phi.
\end{equation}
Through thermodynamic relations and the continuity equation one can obtain $\delta p=-\rho s^{2}{\nabla}^{2}\phi$, 
where $s$ is the adiabatic sound velocity \cite{HYDR}. At the typical frequency $\omega\sim s/a\sim 10^{11}~{\rm s}^{-1}$ 
($a\sim 1~\mu{\rm m}$ is a particle radius), involved into the problem, the dissipative term in Eq.~(\ref{9}) is small and one can write
\begin{equation}
\label{10}
\frac{{\partial}^{2}\phi}{\partial t^{2}}-s^{2}{\nabla}^{2}\phi =0.
\end{equation}
According to the limit of a small friction, the boundary condition for the normal derivative ${\nabla}_{n}\phi=0$ to 
Eq.~(\ref{10}) corresponds to the Euler fluid \cite{HYDR}. From general point of view, the free energy of a system of harmonic 
oscillators does not depend on friction in the thermal limit.

In an electrolyte the dispersion law of sound waves can be approximated as
\begin{equation}
\label{11}
\omega^{2}(\vec q)=s^{2}q^{2}+\frac{s^{2}}{\lambda^{2}_{D}},
\end{equation}
where $\lambda_{D}$ is the Debye screening length. Let us consider first the case of two infinite parallel walls separated by the distance 
$R$. The free energy per unit area of the system is expressed as a sum of energies of independent oscillators according to general 
rules of statistical physics
\begin{equation}
\label{12}
F=T\int\frac{d^2 k}{(2\pi)^2}\sum^{\infty}_{n=1}\ln\left[\omega\left(\vec k,\frac{\pi n}{R}\right)\right].
\end{equation}
Performing the same steps as in Ref.~\cite{IVLEV}, we obtain the interaction mediated by compression fluctuations in the
form 
\begin{equation}
\label{13}
U_{com}=\frac{T}{32\pi R^{2}}\int^{\infty}_{0}dz\ln\left[1-\exp\left(-\sqrt{z+4R^{2}/\lambda^{2}_{D}}\right)\right]
\end{equation}
In the limiting cases Eq.~(\ref{13}) reads
\begin{equation}
\label{14}
U_{com}=
\begin{cases}
-\zeta(3)/16\pi R^{2},&R\ll\lambda_{D}\\
-\exp(-2R/\lambda_{D})/8\pi R\lambda_{D},&\lambda_{D}\ll R,
\end{cases}
\end{equation}
where $\zeta(3)\simeq 1.202$ is the Riemann zeta function. At a large $R$ the interaction (\ref{14}) is screened on the length 
$\lambda_{D}/2$. A possibility of interaction, mediated by non-electromagnetic fluctuations, has been proposed by Dzyaloshinskii, Lifshitz, 
and Pitaevskii \cite{DZYAL}. The first formula (\ref{14}) was obtained in Ref.~\cite{IVLEV}. It is similar to the result of Ref.~\cite{NIN1}
for electromagnetic fluctuations and perfectly conducting planes. 

The fluctuation interaction (\ref{13}) depends on the Debye screening length. This happens due to summation over all wave vectors in the free
energy. This results in its dependence on density of states which, in turn, depends on a form of the spectrum. In our case the spectrum
$\omega=s(k^{2}+\lambda^{-2}_{D})^{1/2}$ brings the $\lambda_{D}$-dependence into the free free energy.

When two objects are not flat but they are close enough interacting by small  parts of their surfaces which are almost flat, one can
derive an interaction potential from the flat approximation (\ref{14}) integrating over the surfaces. For example, for a particle close
to a flat wall when the center-to-wall distance $h$ is smaller than the particle radius $a$, the interaction can be calculated as in 
Ref.~\cite{IVLEV} and has the form 
\begin{equation}
\label{15}
U_{com}=-\frac{\zeta(3)}{8}\,T\frac{a}{h-a},\hspace{1cm}(h-a)\ll a,\lambda_{D}.
\end{equation}
The analogous result for two spheres with the center-to-center distance $R$ is 
\begin{equation}
\label{15a}
U_{com}=-\frac{\zeta(3)}{16}\,T\frac{a}{R-2a},\hspace{1cm}(R-2a)\ll a,\lambda_{D}.
\end{equation}
Equations (\ref{15}) and (\ref{15a}) hold in a non-electrolytic fluid or in an electrolyte with a sufficiently large screening length.
\section{WHICH MECHANISM SURVIVES IN AN ELECTROLYTE}
In an electrolyte the interaction $U_{com}$, mediated by plasmons, is strongly screened (\ref{14}). The van der Waals interaction $u_{vdW}$, 
mediated by photons of a frequency in the visible range, is not sensitive to the plasmon effects. The mechanism of variable 
mass is connected solely with incompressible fluid fluctuations. Therefore, only the conventional van der Waals interaction $u_{vdW}$
and variable mass mechanisms $I$ can survive in an electrolyte. The interaction due to variable mass is considered in the following 
sections.
\section{VARIABLE MASS MECHANISM}
\label{mech}
For the purpose of illustration of the variable mass mechanism one can consider a simple mechanical analogy. Suppose a classical 
non-dissipative particle of the total energy $E$ moves in the harmonic potential  $\alpha x^{2}$. When the particle mass is a constant the 
mean displacement $\langle x\rangle = 0 $ in the harmonic potential. But in the case of a variable mass $m(x)$ the mean displacement 
$\langle x\rangle\neq 0$. This is due to that a particle velocity is smaller at a region of a larger mass just to keep the total energy to
be constant. According to that, the particle spends more time at the region of a larger mass. This is equivalent to the certain effective 
attraction $I(x)$ to a region of a larger mass. As shown in Ref.~\cite{IVLEV}, for a slow varying $m(x)$ the total effective potential 
becomes $\alpha x^{2}-(E/2)\ln m(x)$.

A real particle with friction participates the Brownian motion characterized by the certain temperature $T$. We briefly repeat here the main
arguments leading to an effective potential $I$ \cite{IVLEV}. The Langevin equation, describing such processes, has the form  
\begin{equation}
\label{16}
m(x)\ddot{x}+\frac{1}{2}\frac{\partial m}{\partial x}\dot{x}^2+\frac{\partial V(x)}{\partial x}+\eta\dot{x}=
{\rm stochastic\hspace{0.2cm}force}.
\end{equation}
Short time fluctuations of the velocity $\dot{x}$ are well separated from the slow drift in an effective potential. Indeed, according to 
fluctuation dissipation theorem, the mean value of the kinetic energy $\langle m\dot{x}^{2}/2\rangle=T/2$ corresponds to the equipartition 
law and is contributed by short time fluctuations related to the infinitely large circle in the plane of complex frequency. Substitution of 
that mean value to Eq.~(\ref{16}) leads to the effective potential $V(x)+I(x)$ \cite{IVLEV}, where
\begin{equation}
\label{17}
I(x)=-\frac{T}{2}\ln m(x).
\end{equation}
The expression (\ref{17}) is an exact result in the thermal limit (no quantum fluctuations) as the equipartition law.

We remind what happens in the conventional case of position-independent masses. In that situation in the thermal limit one can put all
frequencies to be zero since they provide only quantum corrections to the partition function. The remaining part of the partition function 
is determined solely by a potential energy and does not depend on velocities. The scenario becomes different for a position-dependent mass.
In this case one can use the following general arguments. The partition function $Z$, which is proportional to the phase volume 
$\Delta p\Delta x$, acquires an additional positional dependence $\Delta p\sim\sqrt{Tm(x)}$ following from the momentum channel even in the 
thermal limit \cite{IVLEV}. The free energy $(-T\ln Z)$ results in the interaction potential (\ref{17}) obtained by the rigorous derivation 
procedure.  

In a multidimensional case the kinetic energy is expressed through the mass tensor $m_{ij}(\vec R)$ as
\begin{equation}
\label{18}
K=\frac{1}{2}\hspace{0.1cm}m_{ij}(\vec R)\dot{R}_{i}\dot{R}_{j}\hspace{1cm}(\omega\rightarrow\infty).
\end{equation}
As shown in Ref.~\cite{IVLEV}, in the multidimensional case the interaction due to variable mass has the form 
\begin{equation}
\label{19}
I(\vec R)=-\frac{T}{2}\ln\left[{\rm det}\,m(\vec R)\right].
\end{equation}
The potential (\ref{19}) has a fluctuation origin and is mediated by fast fluctuations of velocity. In terms of coordinates, when the form 
(\ref{18}) is diagonalized, ${\rm det}\,m$ becomes a product of principal values and the interaction (\ref{19}) is reduced to a sum of terms
related to principal coordinates. 
\section{INTERACTION MEDIATED BY THERMAL FLUCTUATIONS OF PARTICLE VELOCITIES}
\label{inter}
To calculate the effective fluctuation potential even for system with a complicated dynamics, one has to find in the high frequency limit 
the mass tensor and to insert it into Eq.~(\ref{19}). When particles in a fluid perform an oscillatory motion with a high frequency, the 
fluid velocity obeys the Euler equation everywhere in the fluid excepting a thing layer close to the particle surfaces \cite{HYDR}. Hence 
for finding the mass tensor in Eqs.~(\ref{18}) - (\ref{19}) one has to solve the Euler equation with the zero boundary condition for a 
normal component of the fluid velocity. For this reason, the mass, corresponding to the high frequency limit of particle dynamics, can be 
called Euler mass. The effective particle masses depend on a fluid mass involved into the motion. The fluid mass depends on an interparticle
distance and therefore the effective particle masses also depend on that distance.

In the case of one particle of a radius $a$ in a bulk fluid the Euler mass tensor has the form~\cite{HYDR}
\begin{equation}
\label{20}
m_{ij}=\frac{4\pi a^{3}}{3}\left(\rho_0 +\frac{\rho}{2}\right)\delta_{ij},
\end{equation}
where $\rho_0$ is a mass density of the particle and $\rho$ is a fluid density. The first term in Eq.~(\ref{20}) is related to the self mass
of the particles and the second one is associated with a fluid motion.
\begin{figure}
\includegraphics[width=5.0cm,height=3.6cm]{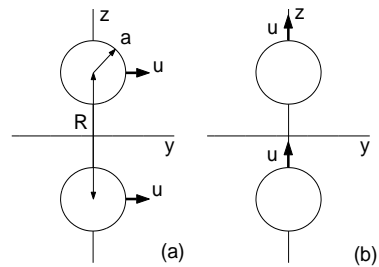}
\caption{\label{fig1} Arrangements of particles with velocities $u$ to calculate in Eq.~(\ref{22}) the mass (a) $M_{x}$ and (b) $M_{z}$.}
\end{figure}
\begin{figure}
\includegraphics[width=5.0cm,height=3.6cm]{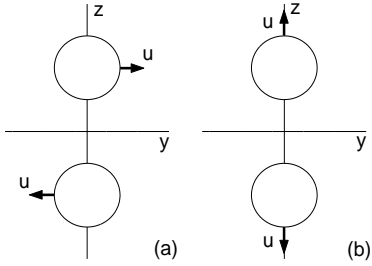}
\caption{\label{fig2} Arrangements of particles with velocities $u$ to calculate in Eq.~(\ref{22}) the mass (a) $m_{x}$ and (b) $m_{z}$.}
\end{figure}
\subsection{Two particles in an infinite fluid}
\label{two}
Now we calculate the Euler mass tensor for two identical particle in a fluid. The fluid velocity, normal to a particle surface, should equal 
a normal particle velocity on its surface. If the particle velocities are $\vec u_{1}$ and $\vec u_{2}$ there are four independent quadratic 
combinations $(\vec u^{2}_{1}+\vec u^{2}_{2})$, $\vec u_{1}\vec u_{2}$, $(\vec R\vec u_{1})(\vec R\vec u_{2})$, and 
$[(\vec R\vec u_{1})^{2}+(\vec R\vec u_{2})^{2}]$. Let us make the velocity transformation 
\begin{equation}
\label{21}
\vec V=\frac{\vec u_{1}+\vec u_{2}}{\sqrt{2}},\hspace{1cm}\vec v=\frac{\vec u_{1}-\vec u_{2}}{\sqrt{2}}.
\end{equation}
In terms of new velocities the kinetic energy can be written in the form
\begin{equation}
\label{22}
K=\frac{1}{2}\sum^{3}_{i=1}\left[M_{i}(\vec R)V^{2}_{i}+m_{i}(\vec R)v^{2}_{i}\right],
\end{equation}
where $i=1,2,3$ correspond to $x,y,z$ respectively. In the expression (\ref{22}) there are four independent masses only
since $M_{x}=M_{y}$ and $m_{x}=m_{y}$. One can separate self masses of the particles $4\pi a^{3}\rho_{0}/3$ from fluid ones using the 
form
\begin{eqnarray}
\label{23}
M_{i}=\frac{4\pi a^{3}}{3}\left[\rho_{0}+\frac{\rho}{2}G_{i}\left(\frac{R}{a}\right)\right],\\
\nonumber
m_{i}=\frac{4\pi a^{3}}{3}\left[\rho_{0}+\frac{\rho}{2}g_{i}\left(\frac{R}{a}\right)\right].
\end{eqnarray}
According to Eq.~(\ref{20}), $G(\infty)=g(\infty)=1$. The easiest way to calculate the masses is to use the method
illustrated in Figs.~\ref{fig1} and \ref{fig2}, where the particle velocities are shown by the arrows. The kinetic energy of the fluid
in Figs.~\ref{fig1}(a) and \ref{fig1}(b) are $(2\pi a^{3}/3)G_{x,z}(R/a)u^{2}$ respectively. Analogously, the fluid kinetic energies in
Figs.~\ref{fig2}(a) and \ref{fig2}(b) are $(2\pi a^{3}/3)g_{x,z}(R/a)u^{2}$. The interaction potential (\ref{19}) takes the form
\begin{eqnarray}
\label{24}
I(R)=-T\bigg[\ln\frac{[2\rho_{0}+\rho G_{x}(R/a)][2\rho_{0}+\rho g_{x}(R/a)]}{(2\rho_{0}+\rho)^{2}}\\
\nonumber
+\frac{1}{2}\ln\frac{[2\rho_{0}+\rho g_{z}(R/a)][2\rho_{0}+\rho G_{z}(R/a)]}{(2\rho_{0}+\rho)^{2}}\bigg].
\end{eqnarray}
The function $I(R)$ tends to zero at $R\rightarrow\infty$. 
\subsection{One particle near an infinite wall}
\label{one}
We  consider here one particle of the radius $a$ placed in a half-space of fluid filled out the volume $z>0$. The center-to-plane (the plane
is at $z=0$) distance is $h$. The boundary condition at the particle surface is an equality of normal components of the fluid and the 
particle velocities. The normal velocity component of the fluid at the flat surface is zero. Obviously, the total kinetic energy is one half
of that calculated in the previous subsection corresponding to Fig.~\ref{fig1}(a) (the $x$ and $y$ components) and Fig.~\ref{fig2}(b) 
which are related to the zero normal velocity of the fluid at $z=0$.

It is easy now to write down the interaction potential using the results of the previous subsection. It reads
\begin{equation}
\label{25}
I(h)=-T\left[\ln\frac{2\rho_{0}+\rho G_{x}(2h/a)}{2\rho_{0}+\rho}
+\frac{1}{2}\ln\frac{2\rho_{0}+\rho g_{z}(2h/a)}{2\rho_{0}+\rho}\right]
\end{equation}
where the functions $G_{x}$ and $g_{z}$ are the same as calculated in the previous subsection. 
\subsection{One particle near two infinite perpendicular walls}
\label{angle}
\begin{figure}
\includegraphics[width=5.5cm,height=3.5cm]{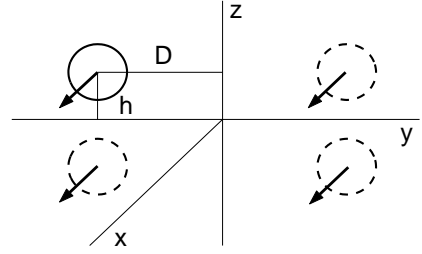}
\caption{\label{fig3} The particle near two perpendicular walls. The three image particles, shown by the dashed curves, are added to 
consider the whole space with zero normal velocities of the fluid at the walls. This particles arrangement contribute to the function 
$f_{1}$ in Eq.~(\ref{26}).}
\end{figure}

\begin{figure}
\includegraphics[width=5.5cm,height=3.7cm]{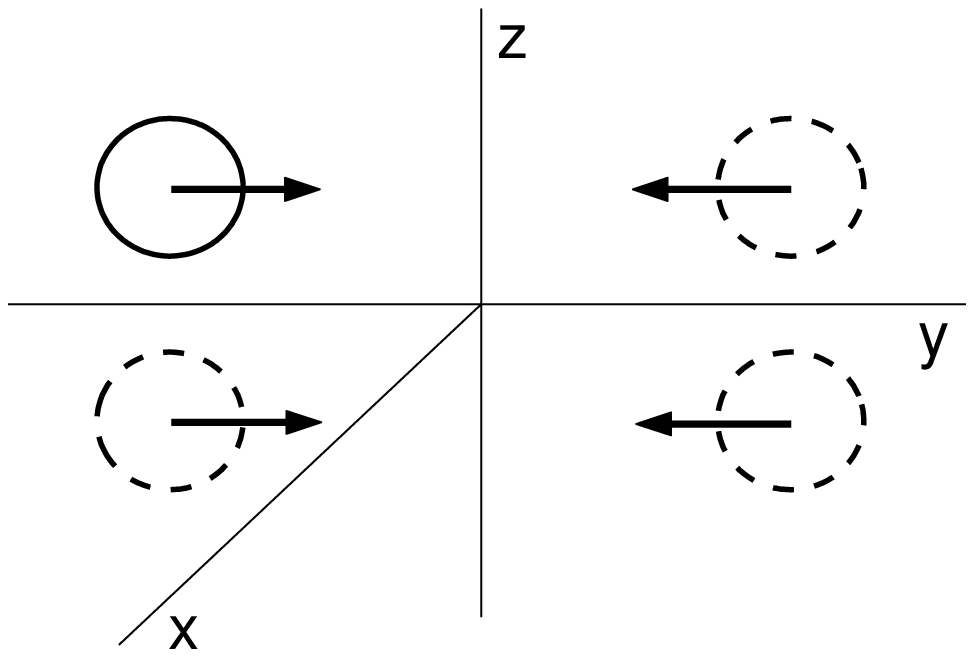}
\caption{\label{fig4} The particles arrangement contribute to the function $f_{2}$ in Eq.~(\ref{26}).}
\end{figure}

\begin{figure}
\includegraphics[width=5.5cm,height=3.5cm]{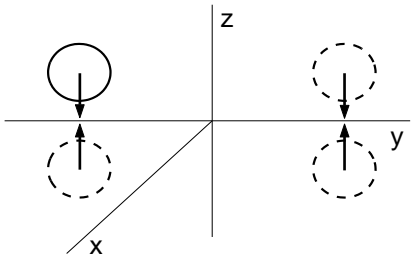}
\caption{\label{fig5} The particles arrangement contribute to the function $f_{3}$ in Eq.~(\ref{26}).}
\end{figure}

This situation is shown in Figs.~\ref{fig3} - \ref{fig5}. The fluid is restricted by the conditions $y<0$ and $z>0$ where the particle is
placed. The other image particles, figured out by dashed curves, serve to obey the boundary conditions in the planes $z=0$ and $y=0$ of zero
normal velocities of the fluid. If to consider the whole space with incorporated image particles the total kinetic energies are 
\begin{equation}
\label{26}
K_{i}=4\frac{4\pi a^{3}}{3}\left[\rho_{0}+\rho f_{i}(h/a,D/a)\right]\frac{u^{2}}{2}
\end{equation}
with $i=1,2,3$ for Figs.~\ref{fig3} - \ref{fig5} respectively where the velocities of all particles are $u$. Analogously to the previous 
cases, the interaction potential is
\begin{equation}
\label{27}
I(h,D)=-\frac{T}{2}\sum^{3}_{i=1}\ln\frac{2\rho_{0}+\rho f_{i}(h/a,D/a)}{2\rho_{0}+\rho}.
\end{equation}
The boundary conditions are of the same type as in Sec.~\ref{two}.

\section{NUMERICAL METHOD TO CALCULATE THE INTERACTION POTENTIAL}
\label{method}
In our case the fluid velocity is $\vec v=\nabla\varphi$, where the potential $\varphi$ obeys the Laplace equation $\nabla^{2}\varphi=0$ with
the boundary conditions specified in Sec.~\ref{inter}
\begin{equation}
\label{28}
\vec n(\vec u-\nabla\varphi)\big|_{S}=0.
\end{equation}
In Eq.~(\ref{28}) $S$ is the total area restricted the fluid including particles, with velocities $\vec u$, and walls with zero velocities.
The unit vector $\vec n$ is perpendicular to the surfaces. The kinetic energy of the fluid is $\int d^{3}r\rho\vec v^{2}/2$. Therefore, the 
total kinetic energy of the system can be written as
\begin{equation}
\label{29}
K=\sum_{i}\left(\frac{\rho_{0}}{2}\,\vec u^{2}_{i}-\frac{\rho}{2}\int_{S_{i}}dS\,\varphi(\vec r)\vec n_{i}\vec u_{i}\right),
\end{equation}
where summation is performed over all particles with the areas $S_{i}$.

The numerical method proposed to calculate Euler masses is based on an iteration procedure. Below we discuss this method for two particles 
in an infinite fluid. The zero approximation is 
\begin{equation}
\label{30}
\varphi_{0}(\vec r)=\psi(\vec r-\vec R/2)+\psi(\vec r+\vec R/2),
\end{equation}
where the harmonic function $\psi$ satisfies the equation
\begin{equation}
\label{31}
\psi(\vec r)=-\frac{1}{4\pi}\int{dS'}\,G(\vec r,\vec r\hspace{0.05cm}')\,
\vec n\hspace{0.05cm}'\,\frac{\partial\psi(\vec r\hspace{0.05cm}')}{\partial\vec r\hspace{0.05cm}'}
\end{equation}
at the exterior of the sphere of the radius $a$ and with the center at $\vec r=0$. At the sphere surface
$\vec n\nabla\psi(\vec r)=u\cos\theta$ and $\theta$ is the angle between $\vec n$ and $\vec u$. The integration in Eq.~(\ref{31}) is 
extended over the sphere surface and Green's function has the form
\begin{equation}
\label{32}
G(\vec r,\vec r\,')=\frac{2}{|\vec r-\vec r\,'|}+
\frac{1}{a}\ln\left(\frac{r-r\cos\gamma}{|\vec r-\vec r\,'|+a-r\cos\gamma}\right),
\end{equation}
where $|\vec r-\vec r\,'|=\left(r^2+a^2-2ar\cos\gamma\right)^{1/2}$ and  
$\cos\gamma=\cos\theta\cos\theta'+\sin\theta\sin\theta'\cos(\phi-\phi')$ \cite{BART}.

The potential $\varphi_{0}(\vec r)$ in the zero approximation (\ref{30}) provides the correct boundary conditions at the particle surfaces 
$S$ in the limit $R\rightarrow\infty$. At a finite $R$ the boundary condition (\ref{28}) is not satisfied by the zero approximation 
(\ref{30}) which requires a modification. One can construct the iteration scheme $\varphi=\varphi_{0}+\varphi_{1}+\varphi_{2}+...$ by means 
of the recursion relation
\begin{eqnarray}
\label{33}
\varphi_{n+1}(\vec r)=-\int_{|\vec r\,'-\vec R/2|=a}\frac{dS'}{4\pi}G(\vec r-\vec R/2,\vec r\,'-\vec R/2)\\
\nonumber
\vec n\,'(\vec u\delta_{no}-\nabla\varphi_{n}(\vec r\,'))\\
\nonumber 
-\int_{|\vec r\,'+\vec R/2|=a}\frac{dS'}{4\pi}G(\vec r+\vec R/2,\vec r\,'+\vec R/2)\\
\nonumber
\vec n\,'(\vec u\delta_{no}-\nabla\varphi_{n}(\vec r\,')),
\end{eqnarray}
where $n=0,1,2,...$. At a large $R$ there is a fast convergence since $\varphi_{n+1}\sim\varphi_{n}a/R$. At each iteration step the boundary
conditions becomes more and more exact with respect the parameter $a/R$.

To calculate $G_{x}=G_{y}$ and $G_{z}$ in Eq.~(\ref{22}) for Euler masses one has to apply the scheme (\ref{33}) to the situations shown
in Fig.~\ref{fig1}(a) and in  Fig.~\ref{fig1}(b). Analogously, $g_{x}=g_{y}$ and $g_{z}$ in Eq.~(\ref{25}) are associated with 
Fig.~\ref{fig2}(a) and Fig.~\ref{fig2}(b). 

By means of the functions $G_{x}$ and $g_{z}$, numerically found by the above method, one can also construct the interaction (\ref{25}) of
a particle and a wall. 

The same method is applicable to calculations of the potential (\ref{27}) in Sec.~\ref{inter} C for one particle near two perpendicular 
walls. Instead of two particles one should take four ones when three of them play a role of images and are shown by the dashed curves in
Figs.~\ref{fig3} - \ref{fig5}. We do not describe an obvious modification of Eq.~(\ref{31}) for that case. 
\section{RESULTS}
\label{disc}
In this chapter we discuss the three different contributions to interaction of colloidal particles, listed in Sec.~\ref{intr}, in various 
geometries: (i) two particles in a bulk fluid, (ii) one particle over a flat surface, and (iii) one particle near two perpendicular planes. 

\begin{figure}
\includegraphics[width=6.5cm,height=4.8cm]{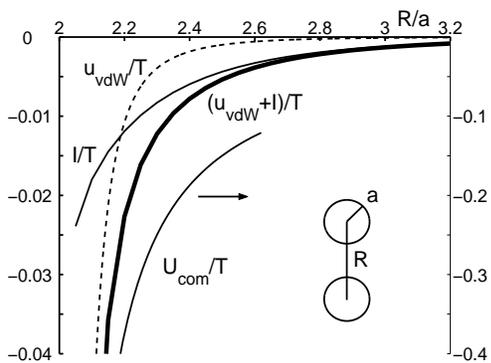}
\caption{\label{fig6} The numerically calculated attraction potentials for two particles in an infinite fluid. $\rho/\rho_{0}=1$.}
\end{figure}

The interaction $U_{com}$, which is mediated by compression fluctuations of the fluid, plays an outstanding role since it is relatively 
large, $U_{com}/T\sim (0.3 - 1.0)$, as follows from Figs.~\ref{fig6} and \ref{fig8}. On the other hand, the interaction $U_{com}$ is strongly
reduced by a finite Debye screening length $\lambda_{D}$. Therefore, this interaction can be observed in electrolytes with a large 
$\lambda_{D}$. In addition to that, surfaces of interacting objects should not be strongly charged to prevent domination of the Coulomb 
repulsion over $U_{com}$. 

In a conventional electrolytes, normally used in experiments, only $u_{vdW}$ (frequencies larger than plasma one) and $I$ (non-compressive 
fluctuations) survive on distances larger than $\lambda_{D}$ where the Coulomb repulsion of charged particles is screened.
\begin{figure}
\includegraphics[width=6.5cm,height=5cm]{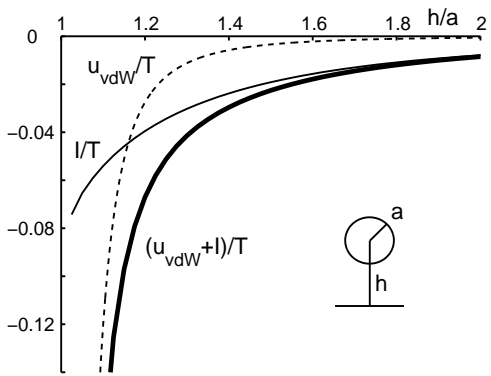}
\caption{\label{fig7} The numerically calculated attraction potentials for one particle and the infinite flat wall. $\rho/\rho_{0}=1$.}
\end{figure}

Fig.~\ref{fig6} relates to the case of two particles in a bulk fluid, Sec.~\ref{inter} A. The interaction potential $I$ (in the units of 
temperature $T$) is plotted by the thin solid curve. The conventional van der Waals interaction (\ref{5}) is indicated by the dashed curve. 
The resulting potential is shown by the thick curve. It can be seen that $U_{com}$ substantially exceeds the above interaction potentials.

Figs.~\ref{fig7} and \ref{fig8} correspond to the case of one particle near the wall, Sec.~\ref{inter} B. As clear from Fig.~\ref{fig8}, 
$U_{com}$ is not small and is of the order of $T$. 

In Fig.~\ref{fig9} the potential $I$ relates to one particle near two perpendicular walls. 
\begin{figure}
\includegraphics[width=6.5cm,height=5.0cm]{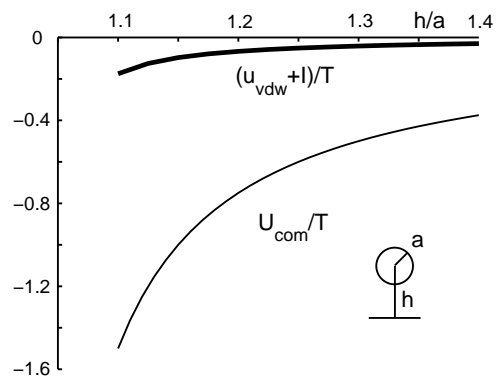}
\caption{\label{fig8} The thick curve is the same as in Fig.~\ref{fig7}. The thin curve represents the interaction (\ref{15}) mediated by 
thermal compression fluctuations of the fluid.}
\end{figure}

\begin{figure}
\includegraphics[width=7.8cm,height=6cm]{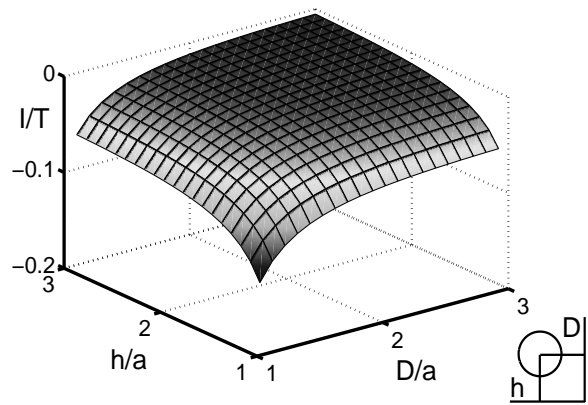}
\caption{\label{fig9} The attractive interaction $I$ in the units of $T$ for one particle and two perpendicular walls. $\rho/\rho_{0}=1$.}
\end{figure}

When the intersurface distance is small the potential $I$ tends to a constant whereas the van der Waals interaction diverges as known.
The results shown in Figs.~\ref{fig6} - \ref{fig9} correspond to $\rho/\rho_{0}=1$. Under increase of this parameter the attraction becomes 
more pronounced. For example, for $\rho/\rho_{0}=13.6$, related to colloidal particles in mercury, the attraction potential becomes roughly 
more than two times larger in amplitude. 
\section{DISCUSSIONS}
Is it possible to experimentally observe an attraction of like-charged colloidal particles separated by micron distance? Let us analyze
different physical situations.
\subsection{No Coulomb effects}
A lack of Coulomb effects means use of an electrolyte with a large Debye length $\lambda_{D}$ and zero charges on all surfaces. In this case 
DLVO repulsion is absent and the interaction mediated by compression fluctuation, $U_{com}$, substantially dominates the conventional van 
der Waals part $u_{vdW}$ and one mediated by particle velocities, $I$. As follows from Fig.~\ref{fig8}, the interaction energy $U_{com}$ of 
a particle of the radius $1~\mu$ and a flat plane, when surface-to-surface distance is $0.15~\mu$, is $1T$. This attraction can be 
experimentally observed since it is relatively large. 
\subsection{Conventional particles in an electrolyte}
For typical electrolytes and polystyrene particles \cite{GREER1} DLVO repulsion is of a short length and the interaction $U_{com}$ is well 
suppressed. In this situation only $u_{vdW}$ and $I$ survive. But the van der Waals interaction $u_{vdW}$ is weak and only $I$ has chances 
to result in an observable attraction. 

The peculiarity of the interaction $I$ is that a particle is attracted stronger to a surface which better geometrically adjusts
its shape. The flat surface in Fig.~\ref{fig7} adjusts the particle better than the counter-curved neighbored particle in Fig.~\ref{fig6}. 
Also the geometry of two walls in Fig.~\ref{fig9} adjust the particle shape better than one wall in Fig.~\ref{fig7}. An attraction of a 
particle to a surface of the same type of curvature is most stronger. 
 
The feature of the interaction $I$ is that it is of the order of $0.05T$ (a particle near a flat wall) and of the order of $0.1T$ (a 
particle near two perpendicular walls) when the intersurface distance is of the order of $0.2a\sim 2000\AA$. At a larger intersurface 
distances the attraction $I$ exceeds the van der Waals part. 

It is promising to take a surface which adjusts the spherical particle better than two perpendicular planes. For example, it can be a 
cylinder with the particle inside. In this case an interaction is a few or even ten times larger than $0.1T$ (two perpendicular planes). The
minimum value of an interaction potential $0.1T$, when $T$ is room temperature, is approximately a border of experimental resolution. So in 
a conventional electrolyte the attraction a micron sized particle to two perpendicular planes or to an interior surface of a cylinder can be
experimentally observed.
\section{CONCLUSIONS}
For like charged colloidal particles two mechanisms of attraction between them survive when the interparticle distance is larger than the 
Debye screening length. One of them is the conventional van der Waals attraction and the second one is the attraction mechanism mediated
by thermal fluctuations of particle position. The latter is related to an effective variable mass (Euler mass) of the particles 
produced by a fluid motion involved. The most stronger attraction potential (up to the value of temperature $T$) corresponds to a situation 
of uncharged particles and a relatively large Debye screening length. In this case the third mechanism of attraction enters the game. It is 
mediated by thermal fluctuations of fluid density. 
\acknowledgments
We thank J. Ruiz-Garcia, C. Bechinger, and M. Kirchbach for valuable discussions.

\end{document}